\documentclass[preprint,authoryear]{elsarticle}
\usepackage{color}
\usepackage{ulem}
\usepackage{deluxetable}
\usepackage{amssymb}
\usepackage{lscape}

\def\hho  {H$_2$O~}

\def\kms  {km~s$^{-1}$}

\def\masy {mas~yr$^{-1}$}

\def\deg  {\ifmmode {^\circ}\else {$^\circ$}\fi}
\def\porm {\ifmmode {\pm}\else {$\pm$}\fi}
\def\chisqpdf {\ifmmode {\chi^2_{\rm pdf}}\else {$\chi^2_{\rm pdf}$}\fi}
\def\chisq    {\ifmmode {\chi^2}\else {$\chi^2$}\fi}

\def\etal {et al.}

\def\vlsr  {\ifmmode {V_{\rm LSR}}\else {$v_{\rm LSR}$}\fi}

\newcommand{\apj}{ApJ}
\newcommand{\aap}{A\&A}
\newcommand{\mnras}{MNRAS}
\newcommand{\pasj}{PASJ}
\newcommand{\araa	}{ARAA}

\journal{New Astronomy}

\begin{document}

\begin{frontmatter}

\title{Analysis of  \hho Masers in Sharpless 269 \\ 
using VERA Archival data
\\ --- Effect of maser structures on astrometric accuracy}
\author{Makoto Miyoshi}
\address{Division of Radio Astronomy, National Astronomical Observatory of Japan,2-21-1 Osawa, Mitaka, Tokyo 181-8588, Japan}
\ead{makoto.miyoshi@nao.ac.jp}
\author{Yoshiharu Asaki}
\address{Institute of Space and Astronautical Science, 3-1-1 Yoshinodai, Chuou, Sagamihara, Kanagawa 229-8510, Japan}
\ead{asaki@vsop.isas.jaxa.jp}
\author{Keiichi Wada}
\ead{wada@astrophysics.jp}
\author{Hiroshi Imai}
\address{Graduate School of Science and Engineering, Kagoshima University,1-21-35, Korimoto, 
Kagoshima, Kagoshima 890-0065, Japan}
\ead{hiroimai@sci.kagoshima-u.ac.jp}
%
\begin{abstract}
 Astrometry using \hho maser sources in star forming regions is expected to be a powerful tool to
study the structures and dynamics of our Galaxy.
Honma et al. (2007) (hereafter H2007) claimed that the annual parallax of Sharpless 269 is determined within an error of 0.008 milliarcsec (mas), 
concluding that S269 is located at 5.3 kpc $\pm$ 0.2 kpc from the sun, and 
its galactrocetnric distance is $R= 13.1$ kpc. 
From the proper motion, they claimed that  the galacto-centric rotational velocity of S269 is 
equal to that of the sun within a 3\% error. 
This small error, however, is hardly understood when taking into account the results of
 other observations and theoretical studies of galactic dynamics. 
We here reanalyzed the VERA archival data using the self calibration method (hybrid mapping), 
and found that clusters of maser features of S269 are distributed in much wider area than
that investigated in H2007. We confirmed that, if we 
make a narrow region image without considering the presence of multiple maser spots,  
and only the phase calibration is applied, we can reproduce the same maser structures in
a maser feature investigated in H2007.  
The distribution extent of maser spots in the feature differs  0.2~mas  from east to west
between our results and H2007.
Moreover, we found that change of relative positions of maser spots in the cluster 
reaches 0.1~mas or larger between observational epochs. 
This suggests that if one simply assumes  the time-dependent, widely distributed 
maser sources as  a stable single point source, 
it could cause errors of  up to 0.1~mas in the annual parallax of S269.
Taking into account the internal motions of maser spot clusters,
the proper motion of S269 cannot be determined precisely. 
We estimated that the peculiar motion of S269 with respect to  a Galactic circular
rotation is $\sim20$~\kms.
{These results imply that the observed kinematics of maser emissions in S269
cannot give a strong constraint on dynamics of the outer part of the Galaxy, 
in contrast to the claim by H2007.}
\end{abstract}



\end{frontmatter}

\section{Introduction}\label{sect:Introduction}

Very long baseline interferometric (VLBI) astrometry of the Galactic maser sources is expected to be a powerful probe for investigation of the structure and kinematics of the Milky Way.
The Very Long Baseline Array (VLBA), the European VLBI Network (EVN), and the Japanese VERA (VLBI Exploration of Radio Astrometry) have been used to measure annual parallaxes and proper motions of star-forming 
regions and red super-giants in spiral arms of the Milky Way \citep[and references therein]{Reid:09, Sato:10}.
\citet{Reid:09} reported that these young sources have large peculiar 
motions (i.e., deviations from circular rotation) as large as 30~\kms. Such large peculiar 
motions are incompatible with the prediction from the conventional theory of 
quasi-stationary spiral arms \citep{LinShu:64, Bertin:96}, but in good agreement with 
recent theoretical high-resolution N-body/hydrodynamical simulations \citep{Baba:09,Wad:01}.
Baba et al. (2009) suggested that spiral arms in the Milky Way are not stationary; in their simulations the arms recurrently
form and vanish. 
Owing to gravitational interactions between the time-dependent spiral 
potential and the ISM,  they showed that the dense gas and star forming regions  have large peculiar 
velocities.

Among the star forming regions whose distances have been measured using 
VLBI, Sharpless 269 (S269) is of special interest. \citet{Honma:07} 
(hereafter H2007) measured the annual parallax and the secular proper motion of S269 
using the VERA, and reported that S269 is located at the 
galactocentric distance, $R$, 
of 13.1~kpc, and its galactocentric rotational velocity is equal to (within 3\%) that 
of the Sun with assumptions of $R_0 = 8$~kpc, and $\Theta_0 = 200$~\kms. 
From these results, they concluded that the flat rotation of the disk of 
the Milky Way extends to 13~kpc from the Galactic center. 
On the contrary,
\citet{Oh:10} observed star forming regions 
AFGL~2789 and IRAS~06058+2138 using the VERA, and concluded that their rotation 
velocities are significantly smaller than the value derived from the assumption 
of the flat rotation. Since the galactocentric distances of these two objects 
are 8.8 and 9.7~kpc, respectively, they suggested that there is a dip in the rotational velocity at around $R \sim 9$~kpc.
 {  If  those objects have large peculiar velocities as suggested by
the theoretical studies, their motions 
may not place strong constraints on the rotation curve.
If S269 has almost the same rotational speed of the sun at $R=13$ kpc  with a small
peculiar motion as suggested by H2007, 
then we should consider how we can reconcile this to other observations and theories.}

H2007 reported that the annual parallax, $\pi$, of S269 is $0.189\pm0.008$~mas, 
which corresponds to $5.28^{+0.24}_{-0.22}$~kpc. 
Given the source distance of 5.28~kpc, the proper motion vector was estimated to be 
($v_{\mathrm{l}}$, $v_{\mathrm{b}}$)
$=$
($-4.60 \pm 0.81$, $-3.72\pm 0.72$)~km~s$^{-1}$. 
The errors in the annual parallax and proper motions of S269 are a factor of 3 to 5 smaller 
than those in recent other VLBI astrometric observations for star forming regions 
\citep[e.\,g.,][]{Sato:10}.
H2007 showed that maser sources of S269 have a simple disk-like 
structure aligned in the east--west direction on a scale of 0.4~mas and a radial velocity 
to the LSR in the range of 19.0 and 20.1~\kms. 
Previous observations, however have suggested 
more complex and time-varying structure in { a wider field} \citep{Lo:73,Genzel:77, White:79, Cesaroni:90,  Migenes:99, Lekht:01a, Lekht:01b}.  
From the \hho maser spectrum shape of the double or triple peaks
around $\vlsr=14~$to$~22~$\kms, two possible structures were proposed by \cite{Lekht:01a}: 
one is an expanding envelope, and the other is an edge-on (Keplerian) disk 
around a protostar. 
\cite{Lekht:01b} reported on a sinusoidal velocity drift at $\vlsr~\sim~20~$\kms, and 
concluded that it is due to turbulent motions of masing clouds because the estimated 
central mass is too small to be a protostar. 
A wide field spatial distribution for the \hho in S269 was shown with 
the VLBI fringe rate mapping technique by \citet{Migenes:99}. They found four velocity components 
at $\vlsr~=16.5, 17.3, 19.4, $and$~20.7$~\kms~spread over 1.3~arcseconds on the sky. 
They also reported that the spatial component $\vlsr~=19.4$~\kms ~was the strongest in their VLBI observation. 
However, the velocity structure of these sources was not studied.  Single-dish 
observation in July 1996 \citep{Lekht:01b} obtained a single peak around 20.3~\kms, 
which probably coincides with the 19.4~\kms ~peak in \citet{Migenes:99}.

{ Three-dimensional velocity estimate of star forming regions may cause big uncertainty in the derived three-dimensional 
motion in the Milky Way because we can often find outflow-like structure in masers which may not 
reflect the motions of the mass center. We have to search the velocity components carefully 
to estimate the motion of the mass centers. In addition, as demonstrated later spatial distributions of maser 
sources also affect the accuracy of VLBI astrometry even for the individual maser spots.
Therefore we have
to investigate the distribution in a wide field for star forming regions.

  In this paper, we focus on the structures of water maser source  in S269 whether it is simple and stable enough to achieve the high accuracy in astrometry using the VERA archival data\footnote{\citet{Rygl:08} and \citet {Rygl:10} reported that they failed to measure the annual parallax for S269's methanol masers with the EVN while 
they obtained the annual parallaxes of five other star forming regions with accuracy as good as $\sim 0.02$~mas.}.}
In section~\ref{sect:observations} we describe the observational specifications.
 Maser emissions in a wide sky area (1.6~arcseconds square) as well as their time 
variations, or relative proper motions,  of the maser spots are shown in section~\ref{sect:results}. We discuss comparison our results with H2007 
and effects of the internal motion of S269 on the Galactic dynamics 
 in section~\ref{sect: discussion} and summarize this study in section~\ref{sect: summary}.

\section{Observations}\label{sect:observations}
As described in H2007, observations have been conducted over one year at 6 epochs on 
Nov~18 in 2004 (Day of Year, or DOY, 2004/323), 
and Jan~26, Mar~14, May~14, Sep~23, and Nov~21 in 2005 
(DOY2005/026, DOY2005/073, DOY2005/134, DOY2005/266 and DOY2005/326). 
The \hho masers at 22~GHz from the S269 region have been observed using 
a 2300~km scale array consisting of four antennas of the VERA
 (Mizusawa, Iriki, Ogasawara, and Ishigaki-jima; see \citealt{Kobayashi:08} in more detail)
 with the left hand circular polarization for almost 8~hours. 
The recording bandwidth for the maser emission was 8~MHz at epochs 1, 4, 5, and 6, 
covering the velocity range of 112~\kms. At epochs 2 and 3, the recording bandwidth 
was 4~MHz to cover the velocity range of 57~\kms. The recorded data was processed 
with the Mitaka FX correlator to produce cross correlated data with the 256 and 512 
frequency channels for epochs~2 and 3, and the others, respectively, so that the 
frequency spacing is 15.625~kHz for all the epochs, corresponding to the velocity 
spacing of 0.21~\kms. 
All the VERA antennas have a dual beam receiving system for phase-referencing \citep{Kobayashi:08},
 and a closely located reference source, J0613+1306, was 
observed simultaneously in the observations. However, we did not carry out data analysis 
of the reference source because our purpose here is concentrated on investigation 
of the spatial and velocity distributions of S269's \hho masers. 

\section{Reduction Methods}\label{sect:Reduction Methods}
\begin{figure*}[ht]
\begin{center}
\includegraphics[scale=0.35]{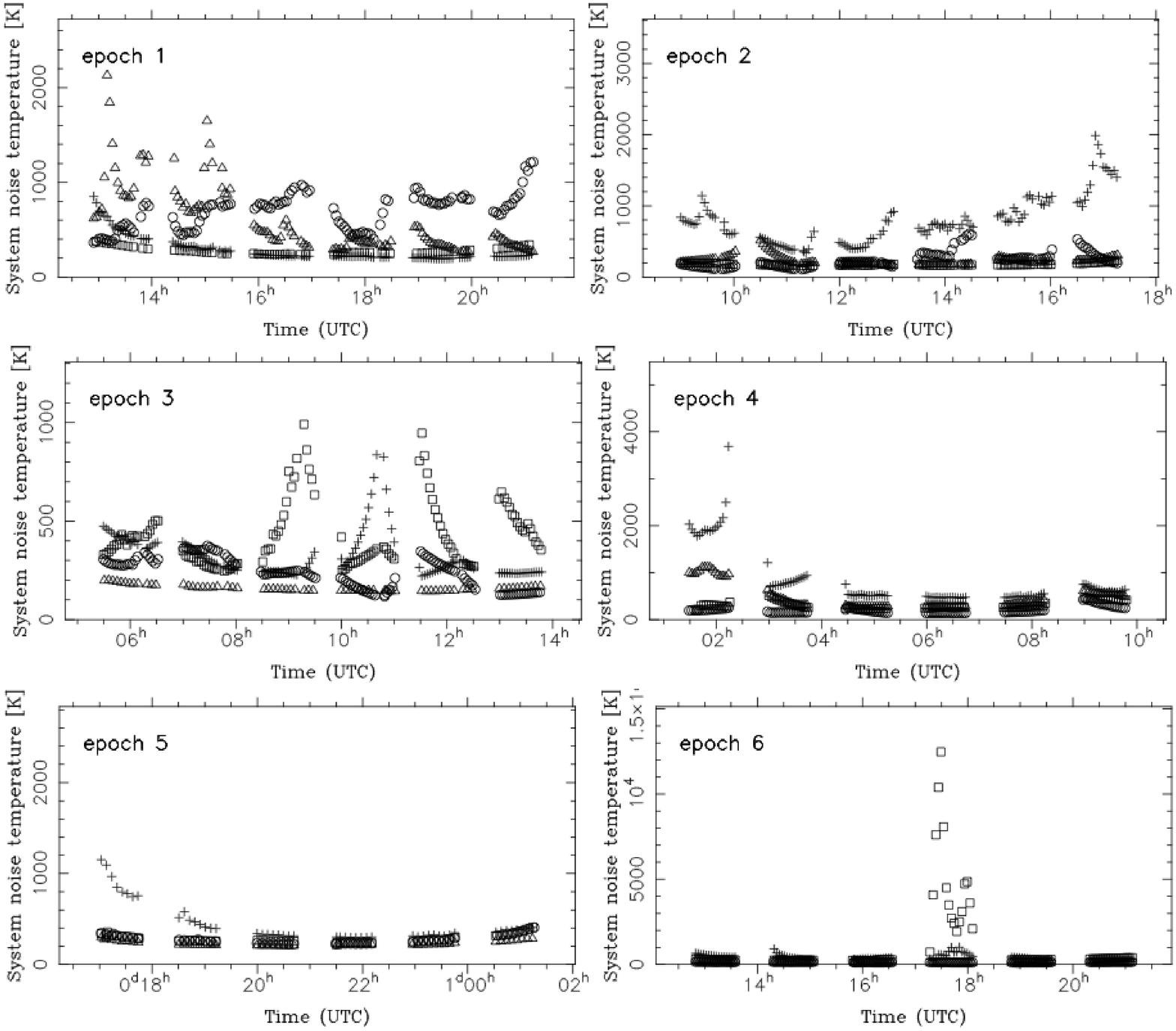}
\caption{System noise temperature for all the observations. The horizontal axis 
is observing time in UTC, and the vertical axis is system noise temperature 
in Kelvin. Circles, triangles, squares, and crosses represent Mizusawa, 
Iriki, Ogasawara, and Ishigaki stations, respectively. }
\label{fig:TsysFig}
\end{center}
\end{figure*}

 In our data reduction, we could not obtain uniform signal-to-noise ratio through all of the
epochs because the atmospheric attenuation were unexpectedly highly variable dependent on observing season in Japan. 
In Table \ref{tbl:tsys} and Figure \ref{fig:TsysFig}, we show the variations of system noise temperatures for all the antennas. 

Basically we followed a standard manner of spectral line VLBI data reductions with AIPS (NRAO) package. However, because of insufficient ($u$,~$v$) coverage of the VERA observations, we found many confusing emission peaks due to the side-lobes coupled with still-not-perfect amplitude calibration. In such a situation, mapping accuracy can depend on the details of data reduction. Here we note the details of our data reduction in order to assure the reproducibility of our mapping results.


\begin{table*}[htbp]
\begin{center}
\caption{
Time average of the system noise temperature of each stations for all the epochs.
}
\begin{tabular}{p{20mm}p{30mm}p{30mm}p{30mm}p{20mm}}
& & & & \\
\hline\hline
Epoch        &  Mizusawa [K]   &  Iriki [K]   & Ogasawara [K]     & Ishigaki [K]    \\
\hline
epoch~1 & 692 & 596 & 280 & 289 \\
epoch~2 & 242 & 230 & 181 & 799 \\
epoch~3 & 240 & 161 & 428 & 328 \\
epoch~4 & 205 & 479 & 311 & 834 \\
epoch~5 & 276 & 234 & 283 & 443 \\
epoch~6 & 137 & 117 & 1050 & 417 \\ 
\hline
\end{tabular}
\label{tbl:tsys}
\end{center}
\end{table*}

\subsection{Visibility Calibrations} \label{sect:vis-cal}
  In order to get reliable images, we must perform calibrations of phase, amplitude, and bandpass characteristics of visibility data.
For visibility-amplitude calibrations, we first performed the task ACCOR with SOLINT=0.1. Using auto-correlation spectra, the task ACCOR corrects amplitude errors in cross-correlation spectra suffered from sampling thresholds. 
 Then we used the task APCAL to generate an amplitude calibration SN table, which includes the information of antenna gain curve (from GC table) and system noise temperatures (from TY table) of each station.
 Furthermore, we applied the amplitude solutions obtained from the self calibration method using the task CALIB.\footnote{In general, measurements of the system noise temperatures and antenna gain parameters are insufficient to calibrate the VLBI data,  because the errors in VLBI data are so large that we cannot rely on conventional calibration and mapping methods often done in connected interferometers. Self calibration in hybrid mapping method provides us powerful solutions for calibrating VLBI data. Today most VLBI maps are obtained after the calibration through hybrid mapping method. For the VERA, due to the insufficient ($u$,~$v$) coverage, it is sometimes difficult to get the optimized solution with hybrid mapping.}\par
 For visibility-delay, rate, and phase calibrations, we used the task FRING with SOLINT=1.0 (SOLSUB=0.1) in order to obtain the clock offset and rate from the strong continuum source, J0530+13 inserted between S269 observing scans. As for fine phase calibrations, we relied on the self calibration solutions from the CALIB in AIPS at the last stage of calibrations (SOLINT=0.1, SOLSUB=0.05).
 The self calibrations were at first performed at the peak frequency channels corresponding to the $\vlsr=19.5$~\kms  (248~ch, 88~ch, 90~ch, 266~ch, 239~ch, and 240~ch in frequency at respective epochs).
 The frequency and velocity resolutions were common through all the observational epochs.
 Although these channels contained strong maser emissions, the maser structures were not a single spot but at least two spots with comparable intensities. The solutions of calibrations from these methods were applied to not only the reference channel but all of the velocity channels.\par
 As for bandpass calibrations we used the task BPASS in AIPS with total power spectra of calibrator continuum sources and got the amplitude bandpass characteristics.
 After these calibrations, we performed corrections of velocity-shift due to diurnal rotation of the earth using the task CVEL in AIPS.
\subsection{How to Make Maps} \label{sect:how-to-map}
\subsubsection{Search for Masing Regions}\label{coarse-mapping}
 To find the whole region of \hho maser emissions, instead of fringe rate mapping,
 we used wide area synthesis imaging with low spatial resolutions.
 We performed synthesis imaging of the $2\times2$~arcseconds area in all velocity channels of four epoch's data (the 1st, the 3rd, the 5th, and the 6th epochs). The $2\times2$~arcseconds area was covered with $4096\times4096$~grids. Namely, each cell size is about 0.5~mas. From these coarse maps, we selected positions for synthesis imaging with fine spatial resolution. We selected positions where the first CLEAN components of the different epochs' data coincided with each other within 0.1~arcsecond. In addition to the positions, we added 0.3 arcseconds square areas around the three positions where strong maser emissions were found (around positions C, D, and E shown in Figure \ref{fig:pmarrow}). We thus selected 42 of $100\times100$~mas areas to be mapped with higher spatial resolution.
\subsubsection{Fine Synthesis Imaging of the Selected 42 Areas}\label{fine-mapping}
We mapped the 42 squares with $4096\times4096$~grids. Each cell size is 24.4$\mu$~arcseconds.
For the synthesis imaging, we used the task IMAGR in AIPS with parameters NITER=3000, GAIN=0.01, and FLUX=15~mJy.
 The selected minimum flux density level of a CLEAN component, FLUX=15~mJy is presumably lower than the array sensitivity. Because the absolute flux density of the data has an uncertainty due to insufficient amplitude calibrations, we used the lower level in order to achieve an adequate subtraction.
\subsection{Measurements of Maser Positions} \label{sect: SAD-selections}
 In order to avoid subjective selections of maser spots, we ran the task SAD in AIPS automatically with its default parameters. We divided the respective $4096\times4096$~grid areas into 25 sub-areas ($4\times4$~mas square), and ran the task SAD in each sub-area and measured maser spot positions.
This selection method partially failed to select some maser positions around strong masers because such regions include not a few numbers of high level peaks due to the side-lobes.
However, we adopted the automatic SAD selection to prioritize objectivity in selecting maser spots.
By the SAD selection we found a lot of peak positions.
The numbers of peaks with flux density $\geq6\sigma$ are given in Table \ref{tab:peakNo}.
 Presumably, side-lobes or not real maser spots mingle among the selected peaks by the SAD method. 
It was quite difficult to select only real maser spots from these maps. To completely avoid selecting peaks that are not real, criteria other than their signal-to-noise ratios (SNR) are required.
\begin{table}[h]
\begin{center}
\begin{tabular}{ccc} \hline \hline
Obs Epoch & Peak Number   & $1\sigma$ Noise Level \\ 
      & $ (>6 \sigma)$ & (Jy/Beam)            \\ \hline
 1    & 238          & $1.37 $ \\ 
 2    & 354          & $4.82 \times 10^{-1}$ \\
 3    & 379          & $4.88 \times 10^{-1}$ \\
 4    & 660          & $1.22 $ \\
 5    & 177          & $1.09 $ \\
 6    & 789          & $5.78 \times 10^{-1}$ \\ \hline
\end{tabular}
\caption{Numbers of the maser emission peaks selected by the SAD method.
  $1\sigma$~noise levels were measured from the $\vlsr=18.5$~\kms channel at the $100\times100$~mas field
  centered at (980~mas, 370~mas) in the map of Figure \ref{fig:pmarrow}. }
\label{tab:peakNo}
\end{center}
\end{table}
%
%
\section{Results}\label{sect:results}

Section~\ref{sect:spectra} shows
the cross power spectra of the \hho maser in the present data analysis. 
Sections~\ref{sect:Emap} and \ref{sect:clusterstructures} show
the spatial distributions of the \hho maser emissions
and its time variation in the individual maser 
spot clusters and the whole area of S269. 
Relative proper motions of the maser clusters are presented in 
section~\ref{sect:relativepropermotions}. 
Here we define a ``maser spot" as the origin of a maser emission
in a single velocity channel map, and a ``maser feature" as a group of maser spots 
with different velocities gathered at a common place. We also use the term
``maser cluster" as a group of maser features with a common motion.
\subsection{
  Cross power spectra of \hho masers in S269
}\label{sect:spectra}

\begin{figure*}[ht]
\begin{flushleft}
\includegraphics[scale=0.7]{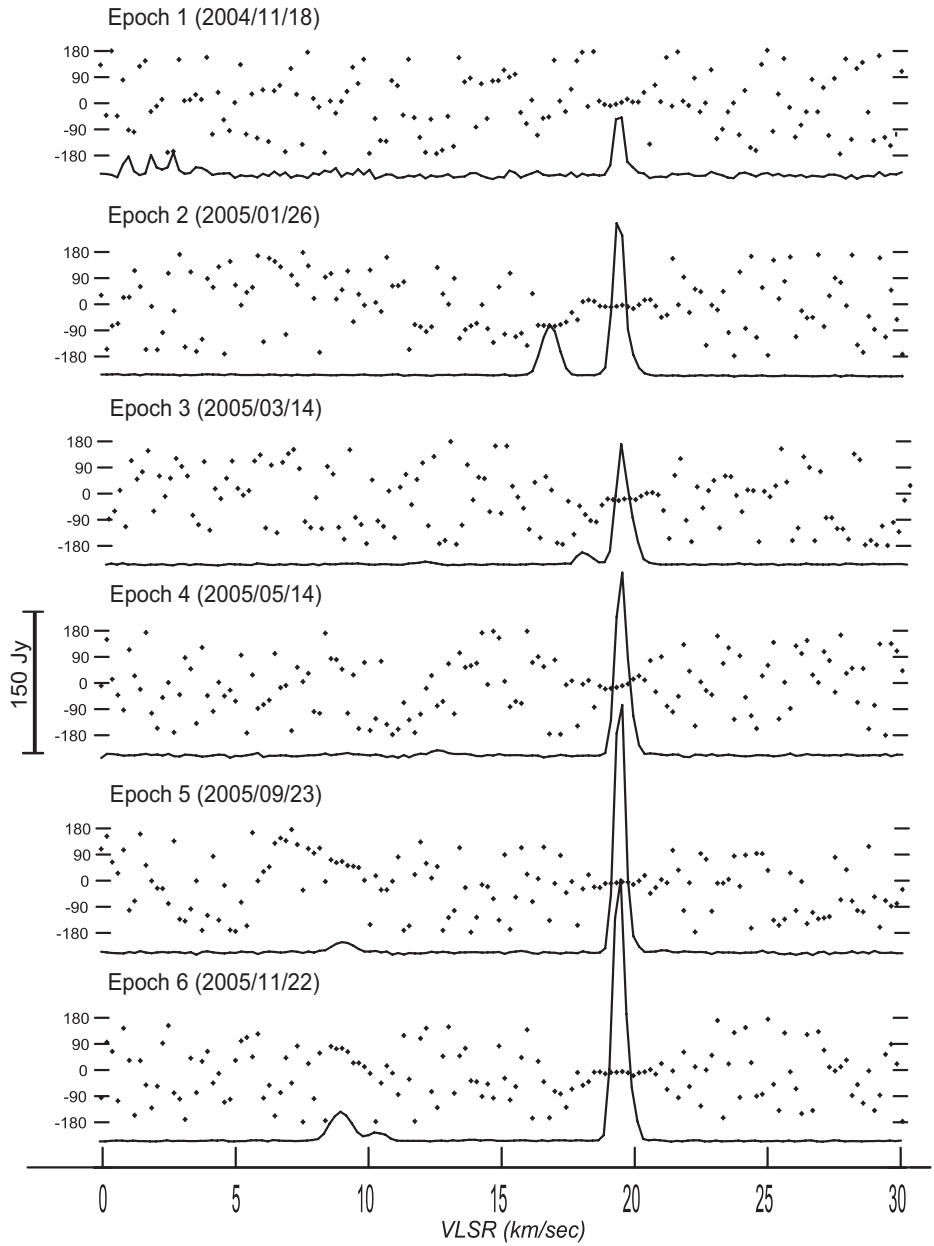}
\end{flushleft}
\caption{
Cross power spectra of the \hho maser emissions in S269 at all the 
epochs obtained from the integration of the entire body of data after 
calibrations for the Mizusawa--Iriki baseline.
Because these amplitudes are obtained from scalar averaging of the cross-power spectra,
the scale has some ambiguity. Several peaks seen between $\vlsr$= 0~to~5~\kms at epoch~1 (top) are
not from \hho maser emissions, but due to artificial radio emissions.
The dots show the phase variations with line of sight velocity channels.
Phase value is shown in degree ranging within $\pm180^{\circ}$.}
\label{fig:cross power spectra}

\end{figure*}

Figure~\ref{fig:cross power spectra} shows the cross power spectra of the \hho maser 
emissions in S269 obtained with the Mizusawa--Iriki baseline (1300~km length) for all 
the six epochs. 
It is important to note that there are complex maser emission peaks in the radial velocity 
range from 8 to 20~\kms, and that the line profile is changed on a time scale 
shorter than one year, as reported by \citet{Lekht:01a}.
The most prominent emission can be seen at 
$\vlsr$~of 19.5~\kms, and the line profile of \hho maser spectrum changes 
significantly in one year. Note also that there are several emission peaks 
in the spectra at 
$\vlsr$~of 17~\kms at epoch~2,  
$\vlsr$~of 18~\kms at epoch~3, and  
$\vlsr$~from 8 to 11~\kms at epochs 5 and 6. 
The signal-to-noise ratio (SNR) of the cross power spectrum at epoch~1 seems worse 
than those at the other epochs. This is mainly because the system noise 
temperatures of these two stations at epoch~1 were unusually a factor of 2 to 5 higher than 
those at other epochs and also because the observing time of S269 was about  half of those in other epochs.

In the cross power spectrum at epoch~1,
there are several peaks between  $\vlsr = 0$ to 5~\kms, but they
are not maser emissions. This is due to strong artificial signals at Iriki station.
We found the artificial signals at 
21.233~GHz (corresponding to $V_{\mathrm{LSR}}$ = 35~km~s$^{-1}$), 
22.235~GHz (corresponding to $V_{\mathrm{LSR}} = 0$~km~s$^{-1}$), and 
22.237~GHz (corresponding to $V_{\mathrm{LSR}}$ = $-22$~km~s$^{-1}$) in sky frequency, 
one of which caused the peaks at $\vlsr$~from 0 to 5\kms.


\subsection{
  Arcsecond scale distribution of \hho masers in S269
}\label{sect:Emap}

Figures~\ref{fig:Emap6-1}, \ref{fig:Emap6-2}, and \ref{fig:Emap6-3} show the spatial 
distribution of the maser spots and position-velocity diagrams for all the six epochs. 
 We identified seven maser groups (A through G). 
All of them show strong time variation. At epoch 1, only group~C was strong (${\rm SNR}>40$). 
C and E were strong at epoch 2, D and E at epoch 3, C and D at epochs 4 to 6. 
Groups~C, E, F and G all lay in $\vlsr$ ranging from 18 to 20~\kms. 
 The maser emissions from $\vlsr~=8$ to 14~\kms ~came from a single group denoted as D.
 The distribution of masers was qualitatively consistent with the map obtained 
 by \citet{Migenes:99}.
They identified four sources at 16.5, 17.3, 19.4 and 20.7 \kms, which roughly coincide with groups A, C, E and G.
They used a fringe-rate mapping method to obtain the positions of the four peaks in their spectrum.
 Therefore, they did not obtain the velocity structures of the individual sources. 
The fact that they did not report the velocity distribution within each maser group does not mean the observed groups 
were single points.

\subsection{Structure of individual maser groups}\label{sect:clusterstructures}

 Figure \ref{fig:spotst} shows the internal maser structures of groups A to G.
The structure of A is taken from epoch 6, that of B is from epoch 4, those 
of C, D, and E are from epoch 3,
and those of F and G are from epoch 6.
 Each maser group consists of one or two features. One feature typically has 
 a spatial size over 1 mas and a velocity range of 2 \kms. Groups D and E consist of 
 two features, while other groups consist of one feature. As shown in the panel for 
group A, the typical beam size is $1.4 \times 0.9$~mas.
 For maser spots in group C, SNR is higher than 100. From the high SNR, 
 we can expect that the error of the relative position is small down to 0.01~mas. For other maser spots, 
 SNR is in a range of 10--20.  { This implies that a possible position error is  
0.1~mas or larger}. 
{ Group C is the brightest among the clusters A-F,
and it should correspond to the maser cluster studied in H2007.}
H2007 detected the maser spots with $\vlsr$ from 19.0 to 20.1~\kms,
  which were distributed over an angular range of 0.4~mas in the east--west direction.
While we detected from group C in wider velocity range of 18.8 to 21.4~\kms ~as shown in Figure~\ref{fig:spotst}.
{ As shown in Fig.  \ref{fig:comparison} (a)}, 
the maser spots in the same velocity range as reported by H2007 
 show a more compact distribution within an angular range of 0.2~mas.
 We discuss the difference between the two maps in section~\ref{sect:comarisonwithH2007}. 

{ The left two panels of Figure \ref{fig:compa} show} the relative positions of the maser spots 
in group~C with respect to the position of the maser spot at 
$\vlsr =$ 19.5~\kms. Positional deviations at epochs~1 and 4 
are larger than those at other epochs: positions at epochs 2, 3, 5, and 6 are consistent 
with one another within the positional accuracy of 0.05~mas.
Judging from the system noise temperatures, the large positional deviations 
at epochs~1 and 4 were caused by the bad atmospheric conditions as noted in section 2.
The plot shows that change of relative positions of maser spots in the cluster 
reaches 0.5~mas between observational epochs (see section 4.1 on possible errors in the annual parallax).

\subsection{Relative Proper Motions of Maser Clusters}\label{sect:relativepropermotions}
 Figure \ref{fig:pm-igs}  shows the proper motions of maser clusters relative to 
the $\vlsr=19.5$~\kms ~spot in group C. 
First, we searched proper motions of maser spots from groups A to G
 (Figure \ref{fig:pmarrow}).
Among the selected peaks, we searched for relative proper motions of maser spots with the following three criteria.

\begin{enumerate}
 \item Peaks which are higher than $20\sigma$ noise level at each epoch.
 \item Peaks whose displacements are less than a corresponding proper motion of  $3.5$~\masy.
 \item Peaks whose velocity shifts are less than $0.5$~\kms.
\end{enumerate}

For peaks in the two fields of groups F and G, 
 where the maser spots were well isolated, 
 we selected peaks  which are higher than $7\sigma$ noise level at each epoch.
 Because these two maser groups show relatively weak emissions but the existence of the groups is certain, 
these masers cannot be neglected in order to investigate the internal motions of the masers in S269 regions.
We found 22 sets of proper motions using the first criterion. 
Using the looser criterion on SNR, we found more 4 sets of proper motions and 1 set of proper motion from the two groups F and G respectively.
In Table \ref{tab:PMT2} we show the detected proper motions of maser spots with these criteria.

After detecting proper motions of maser spots, 
we combined all maser spots data showing common proper motions in order to obtain independent motions.
 First of all, from the proper motions data set, we omitted data using positions of epoch 1 and 4 
as these two epochs were under fairly bad atmospheric conditions. 
We combined and averaged proper motions and velocities of maser spot data whose velocity difference was within 1~\kms~in  order to  derive independent motions of maser clusters.

{   Table \ref{tab:PMT} gives the parameters of derived proper motions of maser clusters.
 Apparently, the maser clusters in groups C, D, E and G form an arc-like structure with the proper motions implying an expanding shell. Since the number of clusters is limited, it is possible to judge that this structure and motion could be just a coincidence.


\normalsize 
\section{
  Discussion
}\label{sect: discussion}

We have re-analyzed the VERA archival data of S269.
We found that
 (a) maser emissions in S269 are distributed over around 1.6 arcseconds,
(b) the maser emissions are found from a wide radial velocity range between 
8 and 20 km $s^{-1}$, 
(c) there are multiple maser groups in the 1.6 $\times$ 1.6 arcsecond area at around the 
radial velocity of 19.5 km s$^{-1}$,  
and 
 (d) the structure of the brightest source (group C) is different from the single source reported in H2007.
 In section \ref{sect:comarisonwithH2007}, 
we discuss the origin of the discrepancy in the obtained maser structure between our result and that of H2007. 
In section  \ref{sect:EforGR},  we discuss the implication of the internal maser motions 
on the constraint on the galactic rotation curve of the Milky Way.

\subsection{Comparison with the map of H2007}\label{sect:comarisonwithH2007}
{ 
In Figure \ref{fig:comparison}, we show the distribution of maser spots of group C and that of H2007, for the same epoch.  Both show velocity gradients in the east-west direction, but H2007's spots in the velocity range of $\vlsr=19.0$ to 20.1~\kms ~are distributed over 0.4~mas, while those in our result have a more compact distribution within 0.2~mas.
 We consider the possibility that the difference between our methods of analysis and those of H2007 caused the difference in the images.
 The primary difference between {the two methods} is that we have made the image from hybrid mapping, while H2007 used the phase-referenced data to determine the absolute positions of the maser spots. This difference of the mapping techniques themselves does not seem to cause
significant differences in the determination of the internal structure.
 Rather, the reasons for the difference  lie in differences in the treatment of the data.

We noted the following two major differences.
 First, H2007 did not find groups E and G, which are in the same velocity range as group C.
 Second, when they performed the analysis of the data, they applied  corrections for the atmospheric delays of stations to the visibility data in order to obtain a reasonable result. They noted: ``To calibrate them, residual zenith delays were estimated as a constant offset that maximizes the coherence of the phase-referenced map. Typical residuals of zenith delay are 1 to 5 cm, but in the worst case (during the summer at Ishigaki-jima station) it was as large as 20 cm".
 
  They maximized the coherence by adjusting the residual atmospheric excess path-length\footnote{In other words, "residual atmospheric delay".},  $L_s$ of individual stations, under the following assumptions:
\begin{enumerate}
\item The excess path-length $\delta L_{S}$ to be adjusted at station S is $\delta L_{S}=L_{S}\times~(\sec~Z_{\rm S269}-\sec~Z_{\rm ref})$, where $Z_{X}$ is the zenith angle of source X at station S and $L_{S}$ is the   residual atmospheric zenith delay at  station S.
\item The residual atmospheric zenith delay $L_S$ is constant during   one observational session.
\item The correct estimate of the residual excess path-length $L_S$ gives the correct map of S269.
\end{enumerate}

 We investigated how these assumptions can change the map. For this purpose we created a map using the method which is effectively equivalent to that used in H2007.
  The difference between the standard imaging method and this is summarized as follows.
 In the standard imaging method one uses fine solutions of both amplitude and phase from self-calibration with an optimized model image, however here we used self-calibration only for phase solutions using a point as the image model.
 Calibration of phase $\phi$ is equivalent to that of excess path-length $L$ due to the equation $\phi=2\pi L/\lambda$ (where $\lambda$ is observing wavelength).
 We assumed that the source in one velocity channel has a single point when we performed the phase self-calibration. We limited the imaging area to a narrow region ($100\times 100$~mas span). 
  If they performed wide-area imaging, they should have found multiple sources as we found. 
(Hereafter we call the map obtained by the above mentioned method as the PPN map: \textit{P}hase-only self-calibration with an one-\textit{P}oint model and \textit{N}arrow field mapping). In Figure~\ref{fig:comparison} we show the PPN map overlaid with the result of H2007. We can see that the PPN map reproduces the main feature of the H2007 map. In particular, the width of the distribution of maser spots is 0.4~mas for both results.
{ We ignored the presence of multiple sources and performed only the phase calibration in the PPN map. This suggests that  the result of H2007 is affected by the imaging area and choice of the calibration method}.

 Figure  \ref{fig:compa} shows the time variation of the relative positions of the maser 
spots in group~C with respect to that at the velocity of 19.5~km~s$^{-1}$ 
revealed by the standard mapping and the PPN method. It is clearly seen 
that the distribution of the spots along the right ascension in the PPN 
is $\sim 0.1$~mas wider than that in the standard mapping method, so that 
the difference between the two as shown in Figure~7 can be reproduced 
for all the epochs. It is worth noting that, assuming that the standard 
mapping method produce true maser emission images better than the PPN 
method because the PPN method uses a single point source model even if 
the structure is complicated, the positional shift seen in the PPN map 
causes an astrometric error, which depends on the source structure. 

Provided that this positional shift have the same direction and the 
same quantity for all the epochs, following astrometric analysis 
to obtain annual parallaxes and proper motions can make a correct 
estimation of the annual parallax. On one hand, if the positional shift 
have a one-year periodical variation, the derived annual parallax 
may have a bias of
 0.1~mas in the worst case. If the positional shift 
appears randomly, the annual parallax error could be between $0-0.1$~mas 
by chance. From our comparison between the standard mapping method and 
the PPN method, and between the PPN method and the map shown in H2007, 
we suspect that H2007 applied a simple source model in their 
analysis which does not include a widely distributed maser emission as 
shown in this report, so that their annual parallax have a hidden 
error of 0.1~mas in the worst case. 
}

\subsection{Effects of the internal motions in S269 on the Galactic dynamics}\label{sect:EforGR}
One of the most important claims in H2007 is that motion and distance to S269 give
a strong constraint on the shape of the outer rotation curve of the Milky Way: the difference of rotational velocities at the Sun and at S269 (which is claimed to be located at 13.1 kpc
away from the Galactic center) is less than 3\%. However, our
reanalysis suggests that we cannot give 
that strong constraint on the outer rotation curve only by
the VERA observation of S269, {if we consider 
the complicated internal structures of maser
clusters and their motions as well as their time variability}.

 We calculated the $UVW$ velocities of the independent maser clusters
 using relative proper motions and radial velocities, assuming that
 H2007 velocity corresponds to that of the $\vlsr=19.5$~\kms ~in the
 maser spot in group C.  { The results are summarized in Table \ref{tbl:UVW2}.} 
 The average UVW of these maser clusters in
 S269 is ($4.4\pm6.1, -16.7\pm18.4, 12.2\pm18.7$ \kms).  Assuming that
 H2007 obtained the correct distance to S269, the rotational velocity at
 13 kpc is $183\pm19.4$~\kms.
%

The rotational velocity at 13 kpc, $183\pm19.4$~\kms,
 is consistent with other measurements.
 Owing to the large
 internal motion of masers in S269, 
 the observed proper motion of S269 with respect to the Local Standard of Rest
cannot give a strong constraint on kinematics of the outer part of the Galaxy, 
in contrast to the claim by H2007. Taking into account the internal motion of
 $\sim20$\kms, the rotational velocity of S269 is consistent with
 other previous observations outside of the solar circle \citep{Sofue:09}}.
\footnote{The annual parallax measurements using the clusters newly 
found in this paper will discussed elsewhere (Asaki et al. in prep.).}
 
\subsection{Difficulties in VLBI astrometry using \hho masers in star forming regions}\label{sect:VLBI-MA}
VLBI astrometry using \hho maser sources has made great progress in decades as a powerful tool to 
investigate structures and dynamics of the Milky Way Galaxy
\citep[e.g.][]{Reid:09, Sato:10}.
However, our result presented here illuminates essential difficulties in this methodology. 
Major intrinsic problems are: 1) \hho masers in star forming regions are not necessarily
located in the `center of mass' of the objects, 2) the masers are not in general
``point '' sources, and their distributions are time-variable, and 
  3) they often show complicated internal motions. 
Additionally, the different results between H2007 and the present work
suggested that 
the data calibration in VLBI observations of star forming regions 
is not straightforward, and still
need improvement. 

%

 One should also note that the shapes of \hho masers are not
 point-like and their spatial extent are often comparable to observational
 beam size. For example, we found that 
 average size of \hho masers in S269 is $0.72\pm 0.50$~mas (HPBW)
based on a Gaussian shape fitting, which is comparable to 
the VERA synthesized beam size at 22 GHz (Figure \ref{fig:spotst}).
The distributions of masers often change between observational epochs,
which brings additional errors on position of the objects.


%
%
%
%
 Moreover, \hho maser sources in a star forming region often show
 internal motions,  i.e. spots apparently moves in terms of the center of mass of the region.
Such internal motions are typically a few tens of km s$^{-1}$,
and therefore it is hard to estimate the kinematic motions of the objects 
in the Galaxy. This could be solved by introducing a model of internal 
motions (e.g. expanding outflows) of maser spots 
\citep[e.g.][]{Imai:00, Asaki:10}.
However, this suggests that a large enough number of spot motions should be measured over
many years, and that the motions are often too complicated to be fitted by 
a simple kinematic model.

 Finally, one should note that the astrometric results of
star forming regions
using the VLBI technique is still carefully considered.
For example, in S269, as shown in section \ref{sect:comarisonwithH2007}, the difference of data calibration and the assumption for measurement may cause up to $0.1 mas$ ambiguity in the astrometric results. 
Although it would be hard to overcome the intrinsic difficulties in
\hho masers of star forming regions, one could reduce errors in astrometry
by searching maser spots as many as possible in a large enough area, and those 
spots should be observed over many years.

\section{Summary}\label{sect: summary}
	
 We have {reanalyzed VERA archival data of S269, and found 
several maser clusters distributed over 1.6 arcseconds 
on the sky.} 
{ The distribution of maser spots in the strongest maser feature differs from that obtained by 
H2007 with $\sim 0.2$ mas. We confirmed that 
this discrepancy is caused by 
unrealistic assumptions and insufficient procedures of analysis.  
If we assume  only a single spot in one velocity channel and perform self phase-only calibration, 
the results of H2007 are reproduced.  }
{ We also found that the relative proper motions and radial motions of multiple maser clusters  are 
quite large}, and therefore the absolute proper motion of S269 does not pose a tight 
constraint on the rotation curve of the Milky Way as was claimed by H2007. 
{ Our analysis also shows that change of relative positions of maser spots in the cluster 
reaches 0.1~mas or larger between observational epochs.
All these results imply that 
the annual parallax of S269 could not be determined within an error of $\sim$ 0.1~mas 
using assumptions of a single point source model for such widely distributed 
maser sources.}

Thus the maser astrometry should be performed carefully 
with consideration about structure 
and the time-variation of maser source. 
The internal motions of maser clusters among the system
 also be considered for Galactic astrometry.
The VLBI astrometric accuracy depends on the way of data analysis.
Thus, the details, as well as the results, should be noted clearly for refinement by future ages.


\newpage
%

\begin{figure*}[htbp]
	\begin{center}
\includegraphics[scale=1.2]{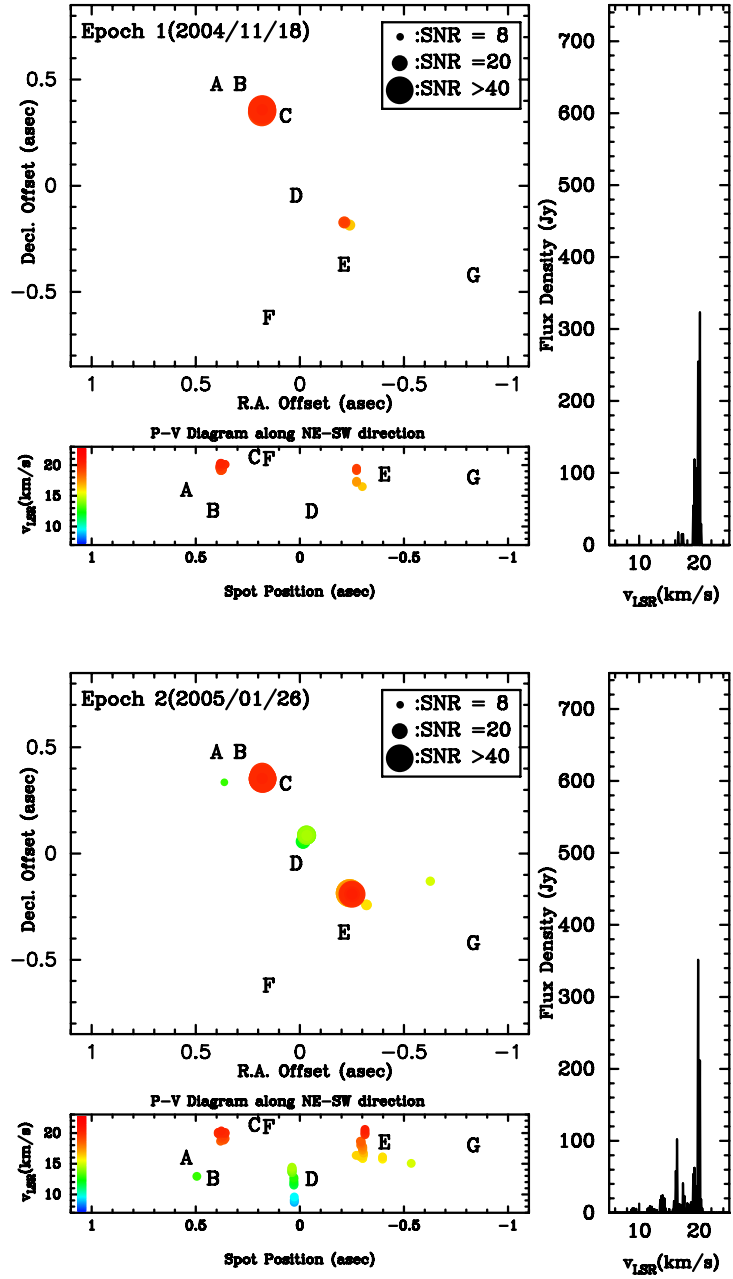}
\caption{Spatial distributions of the \hho maser groups of A to G in S269.
The distributions at epochs 1 \& 2.
Panel top left: the distribution of the maser spots with SNR$\geq 8$.
 In group F, there is no maser spot with SNR$\geq 8$. 
Peaks around group C are not plotted because they are mostly side-lobes.
Right panel: spectra composed from the plotted maser spots.
Bottom left panel: position-velocity diagrams of the plotted maser spots 
cutting along the northeast-southwest.}
\label{fig:Emap6-1}
	\end{center}
\end{figure*}

\begin{figure*}[htbp]
	\begin{center}
\includegraphics[scale=1.2]{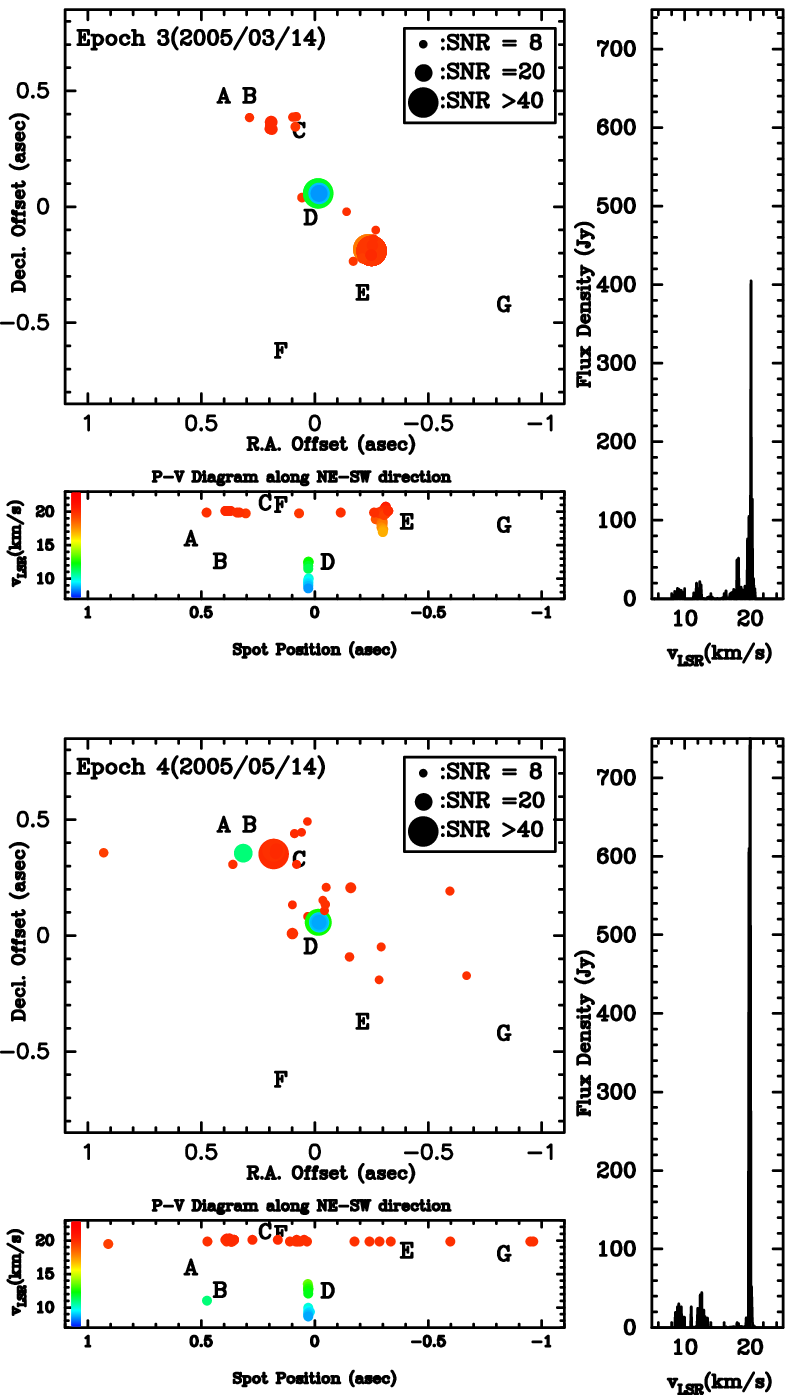}
\caption{Continued. The distributions at epochs 3 \& 4.}
\label{fig:Emap6-2}
	\end{center}
\end{figure*}

\begin{figure*}[htbp]
	\begin{center}
\includegraphics[scale=1.2]{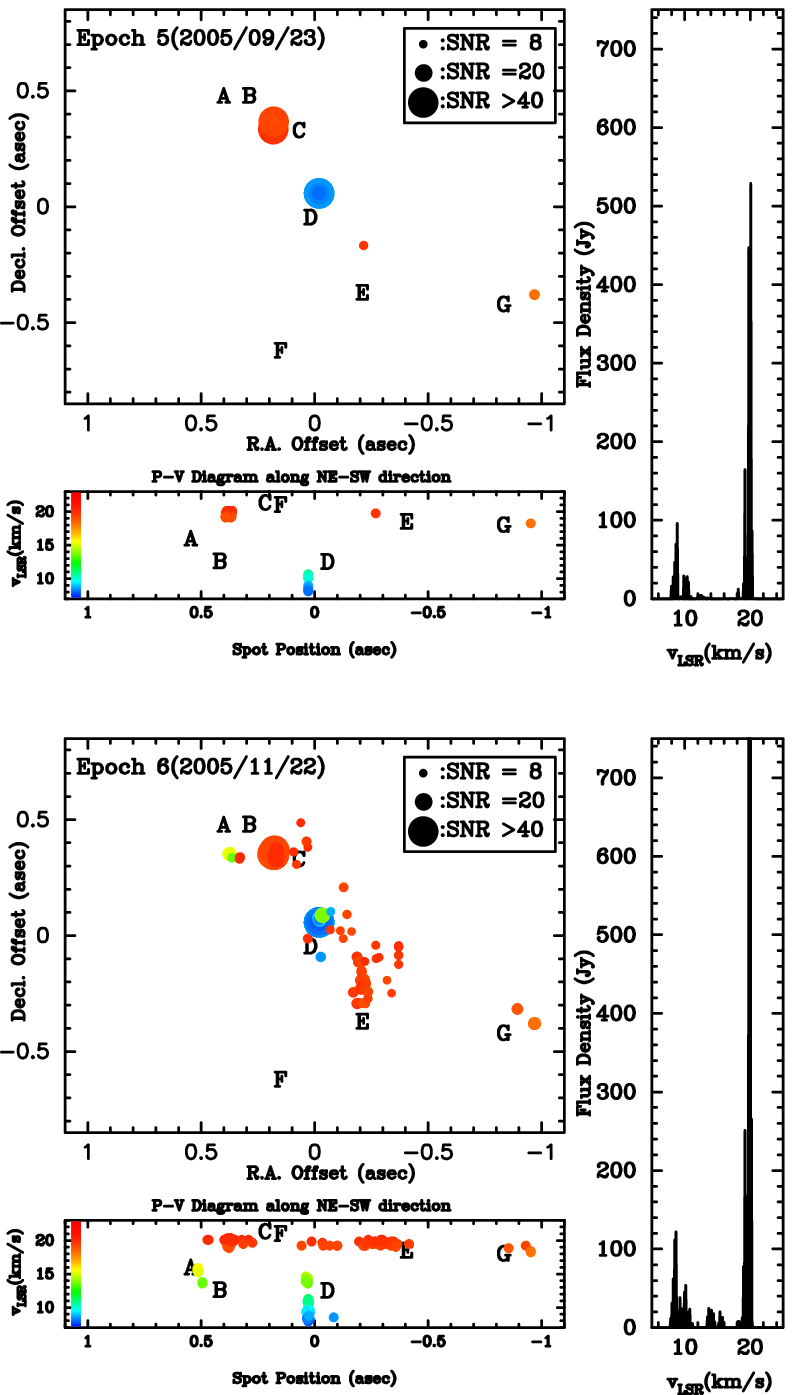}
\caption{Continued. The distributions at epochs 5 \& 6.}
\label{fig:Emap6-3}
	\end{center}
\end{figure*}

\begin{figure*}[htbp]
\begin{center}
\includegraphics[scale=0.15]{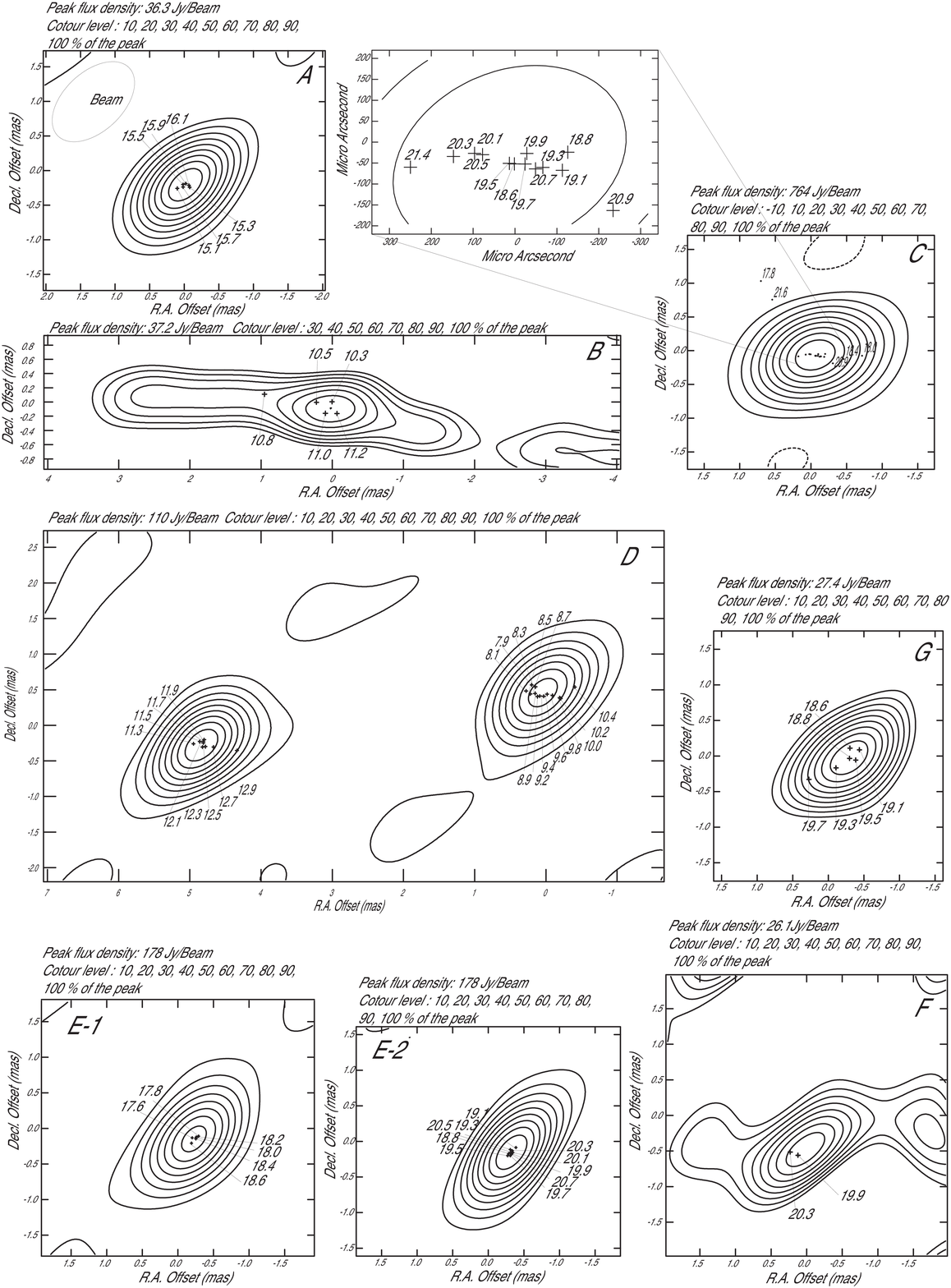}
\caption{
Velocity structures of the respective maser groups in S269.
Velocity integrated maser features are expressed with contour maps.
 Contours are drawn at each 10~\% of the peak intensity.
 Positions of velocity components are plotted by cross marks, denoting numbers 
 indicating the $\vlsr$ velocities.
The structure of A is from epoch 6, that of B is from epoch 4, those of
 C, D, \&E are from epoch 3, and those of F \& G are from epoch 6.
 The typical restoring beam shape ($1.4\times0.9$~mas) is shown at top left in panel A.}
\label{fig:spotst}
\end{center}
\end{figure*}

\begin{figure*}[htbp]
\begin{center}
\includegraphics[scale=0.8]{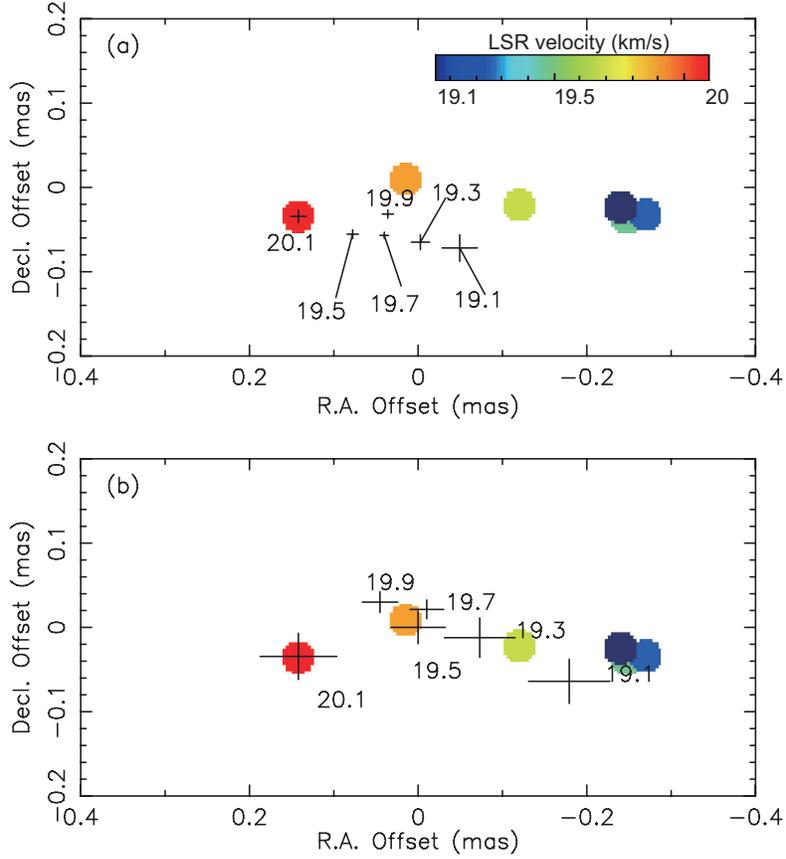}
\caption{
Comparison of the maser spot distribution with that of H2007.
(a) Comparison of our results and that of H2007.
The crosses represent the position of the each maser spot in group C 
obtained from data analysis using the hybrid mapping method.
From our map, we detected maser spots for a wider velocity range of 18.8 to 21.4~\kms 
in the maser group C, but here for comparison we plot the spots limited in the range from $\vlsr=19.1$~to$~20.1$~\kms.
The filled circles are results reported by H2007, and the color represents 
the radial velocity. Two maps are aligned at the spot position of $\vlsr=20.1$~\kms.
(b) Comparison of the results from the PPN method and that of H2007.
The crosses show the position of velocity component $\vlsr=19.1$~to$~20.1$~\kms from 
our reproduction image by the PPN method.}
\label{fig:comparison}
\end{center}
\end{figure*}

\begin{figure*}[htbp]
\begin{center}
\includegraphics[scale=0.7]{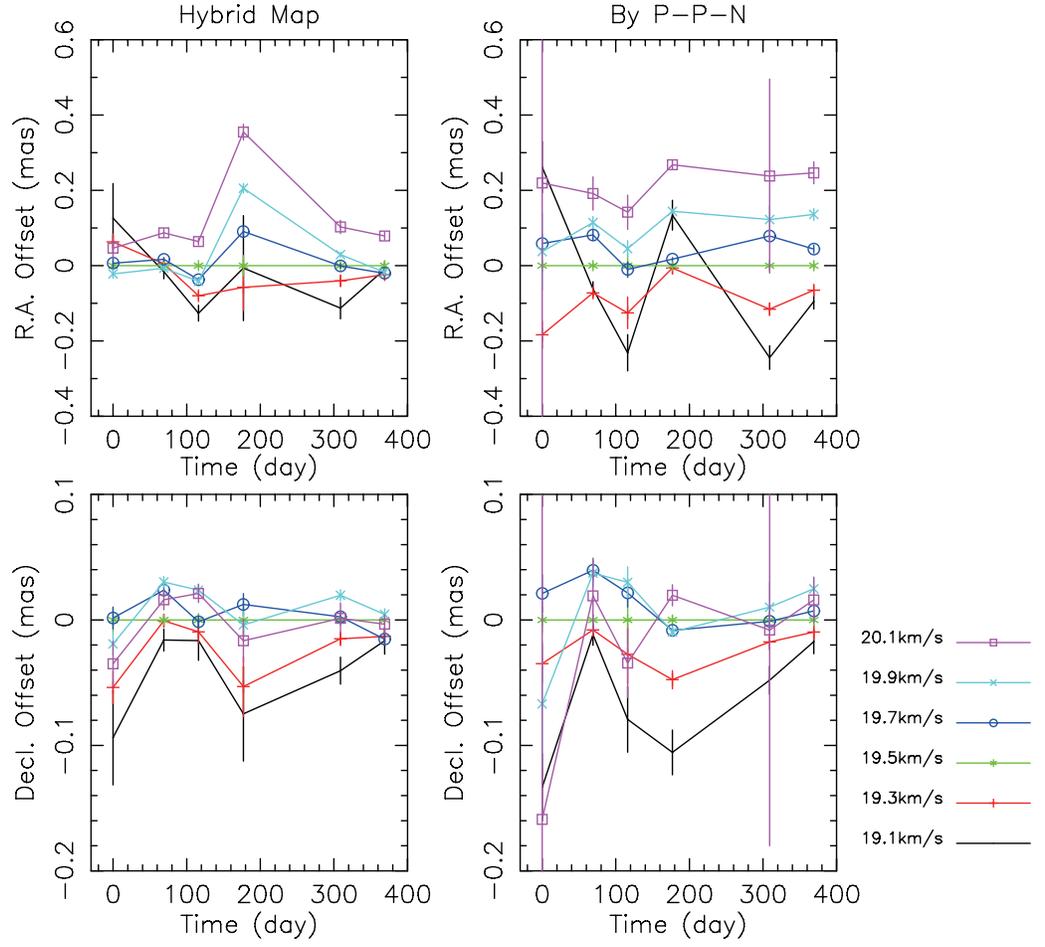}
\caption{
Positional offsets of the maser spots involved in group C with respect to 
the position of the maser spot with $\vlsr = 19.5~$\kms.
Comparisons of our results from hybrid mapping and those from the P-P-N method: 
Left panels are from ours while right panels are from the P-P-N method.
Top panels are for R. A. offsets while bottoms are for declination offsets.}
\label{fig:compa}
\end{center}
\end{figure*}

\begin{figure*}[htbp]
\begin{center}
\includegraphics[scale=0.6]{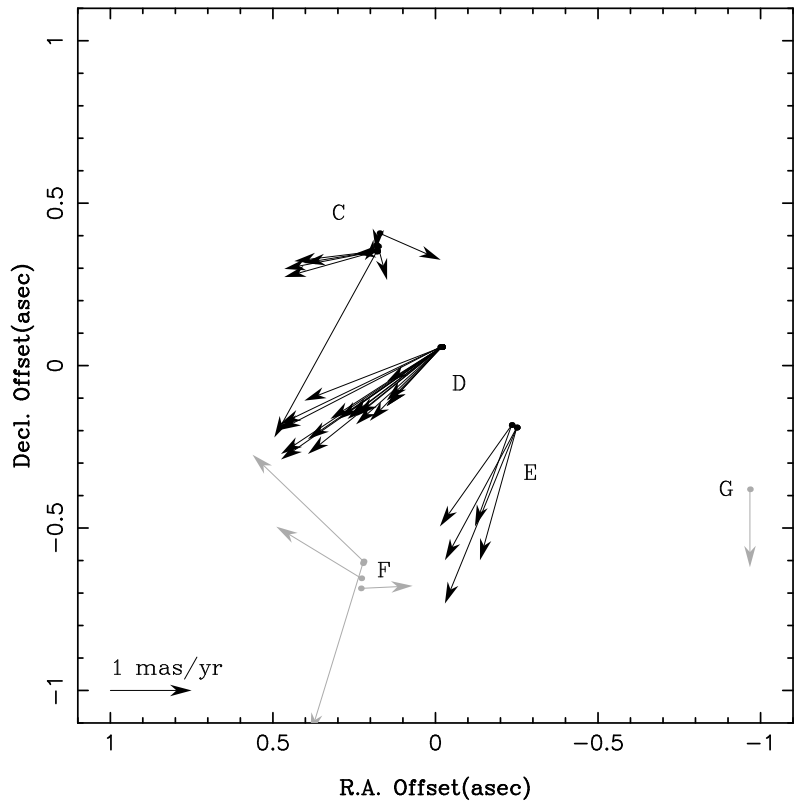}
\caption{
Relative proper motions of the \hho maser spots in S269 with respect to the 
maser spot position with $\vlsr$~of 19.5~\kms.
The proper motions shown by black arrows are detection with SNR$\ge20$ from 
groups C, D, and E. Those shown by gray arrows are detection's with $SNR \ge7$ 
from the weaker groups  F (220 mas, -600mas) and G (-970 mas, 380 mas). }
\label{fig:pmarrow}
\end{center}
\end{figure*}

\begin{figure*}[htbp]
\begin{center}
\includegraphics[scale=0.6]{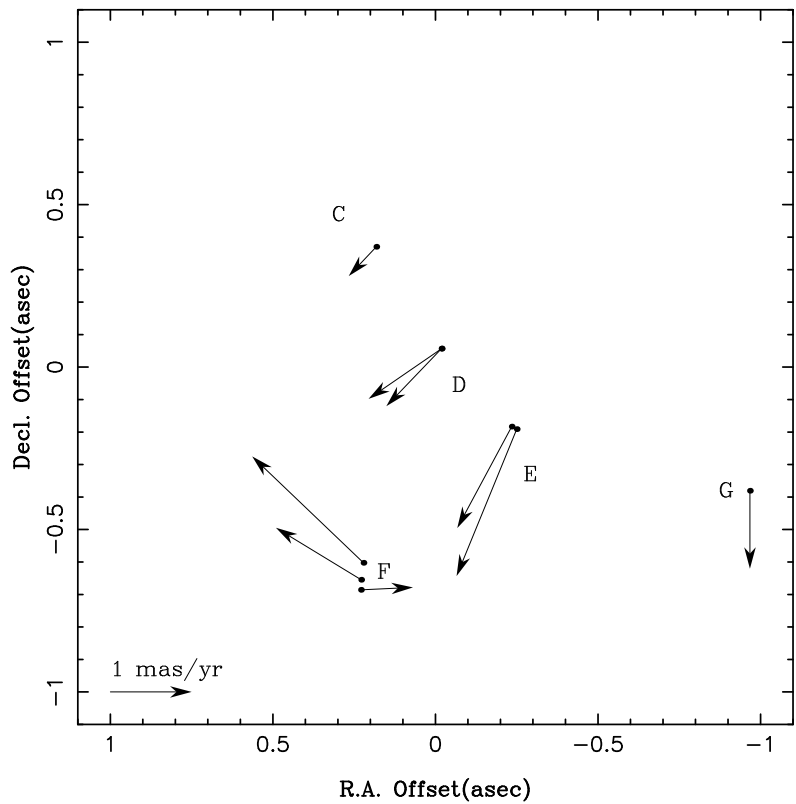}
\caption{
Relative proper motions of the independent 9 \hho maser clusters in S269.
The maps of the 6 observational epochs are aligned using the strong maser spot 
position in the maser group C at $\vlsr=19.5$~\kms.}
\label{fig:pm-igs}
\end{center}
\end{figure*}


\newpage

\begin{landscape}

\begin{deluxetable}{rlrrrrrrrrrrl}
\tablecolumns{13} \tablewidth{0pc}
\tablecaption{Relative Proper Motions of Maser Spots. 
}
\tablehead {
\multicolumn{1}{c}{\footnotesize{No}}&&\footnotesize{Time} &\footnotesize{Velocity}&\multicolumn{2}{c}{\footnotesize{Relative Position}}&\multicolumn{2}{c}{\footnotesize{Intensity}}&\footnotesize{Size}&\multicolumn{2}{c}{\footnotesize{Proper Motion}}
&\footnotesize{Overlap}\\
 &&& \footnotesize{\vlsr} & \footnotesize{$\alpha$} & \footnotesize{$\delta$} &&&\scriptsize{HPBW} &
 \footnotesize{$\mu$} & \footnotesize{$PA$} &  \\
 &&\scriptsize{(day)} &\scriptsize{(\kms)}&\scriptsize{(mas)}&\scriptsize{(mas)}&\scriptsize{(Jy)}&
  \scriptsize{\it(S.N.R.)} &\scriptsize{(mas)}&\scriptsize{(mas~yr$^{-1}$)}&\scriptsize{($^{\circ}$)}&
}
\startdata
~1...1&$^{D1}$&309&~8.3&~-18.98&~~57.14&278&~25.5&0.58&~~~~~&~~~~&~\\ 
~~...2&$^{D1}$&369&~8.3&~-18.87&~~57.04&352&~60.9&0.50&~0.94&~-44&~\\ 
~2...1&$^{D1}$&309&~8.5&~-18.98&~~57.14&465&~42.6&0.53&~~~~~&~~~~&~\\ 
~~...2&$^{D1}$&369&~8.5&~-18.87&~~57.02&626&108.3&0.37&~0.99&~-47&~\\ 
~3...1&$^{D1}$&309&~8.7&~-19.04&~~57.14&696&~63.9&0.31&~~~~~&~~~~&~\\ 
~~...2&$^{D1}$&369&~8.7&~-18.93&~~57.02&977&169.0&0.13&~0.99&~-47&a\\ 
~4...1&$^{~~}$&116&~8.9&~-19.65&~~57.61&125&~25.6&0.70&~~~~~&~~~~&~\\ 
~~...2&$^{~~}$&177&~8.9&~-19.38&~~57.42&252&~20.6&0.59&~2.00&~-35&~\\ 
~~...3&$^{D1}$&309&~8.9&~-19.04&~~57.14&934&~85.7&0.15&~1.46&~-37&b\\ 
~~...4&$^{D1}$&369&~8.7&~-18.93&~~57.02&977&169.0&0.13&~1.35&~-39&a\\ 
~5...1&$^{~~}$&116&~9.2&~-19.71&~~57.61&116&~23.7&0.70&~~~~~&~~~~&~\\ 
~~...2&$^{~~}$&177&~9.1&~-19.43&~~57.39&292&~23.9&0.50&~2.11&~-39&c\\ 
~~...3&$^{D1}$&309&~8.9&~-19.04&~~57.14&934&~85.7&0.15&~1.55&~-35&b\\ 
~~...4&$^{D1}$&369&~8.7&~-18.93&~~57.02&977&169.0&0.13&~1.41&~-37&a\\ 
~6...1&$^{~~}$&116&~9.4&~-19.76&~~57.61&109&~22.4&0.70&~~~~~&~~~~&~\\ 
~~...2&$^{~~}$&177&~9.1&~-19.43&~~57.39&292&~23.9&0.50&~2.38&~-34&c\\ 
~~...3&$^{~~}$&309&~8.9&~-19.04&~~57.14&934&~85.7&0.15&~1.63&~-33&b\\ 
~7...1&$^{~~}$&116&~9.6&~-19.82&~~57.64&101&~20.6&0.64&~~~~~&~~~~&~\\ 
~~...2&$^{~~}$&177&~9.5&~-19.49&~~57.40&255&~20.9&0.52&~2.42&~-35&~\\ 
~~...3&$^{D2}$&309&10.0&~-19.26&~~57.13&277&~25.4&0.28&~1.42&~-42&~\\ 
~~...4&$^{D2}$&369&10.0&~-19.20&~~57.01&451&~78.0&0.36&~1.27&~-46&~\\ 
~8...1&$^{D2}$&309&10.2&~-19.37&~~57.08&263&~24.2&0.67&~~~~~&~~~~&~\\ 
~~...2&$^{D2}$&369&10.2&~-19.26&~~57.01&512&~88.5&0.44&~0.80&~-32&d\\ 
~9...1&$^{D2}$&309&10.4&~-19.54&~~57.11&283&~25.9&0.48&~~~~~&~~~~&~\\ 
~~...2&$^{D2}$&369&10.2&~-19.26&~~57.01&512&~88.5&0.44&~1.82&~-21&d\\ 
10...1&$^{~~}$&116&12.3&~-14.98&~~56.92&218&~44.7&0.73&~~~~~&~~~~&~\\ 
~~...2&$^{~~}$&177&12.5&~-14.65&~~56.74&418&~34.2&0.60&~2.23&~-27&e\\ 
11...1&$^{~~}$&116&12.5&~-14.98&~~56.90&130&~26.6&0.69&~~~~~&~~~~&~\\ 
~~...2&$^{~~}$&177&12.5&~-14.65&~~56.74&418&~34.2&0.60&~2.20&~-26&e\\ 
12...1&$^{E1}$&~69&18.2&-234.74&-182.96&112&~23.2&0.51&~~~~~&~~~~&~\\ 
~~...2&$^{E1}$&116&18.2&-234.68&-183.12&435&~89.1&0.65&~1.33&~-70&h\\ 
13...1&$^{E1}$&~69&18.4&-234.79&-182.96&110&~22.8&0.31&~~~~~&~~~~&~\\ 
~~...2&$^{E1}$&116&18.2&-234.68&-183.12&435&~89.1&0.65&~1.53&~-55&h\\ 
14...1&$^{~~}$&~~0&19.3&~180.82&~350.61&321&~23.4&2.33&~~~~~&~~~~&~\\ 
~~...2&$^{~~}$&369&19.2&~181.10&~350.59&709&122.7&1.39&~0.28&~~-6&~\\ 
15...1&$^{~~}$&~~0&19.5&~176.43&~367.39&400&~29.2&0.60&~~~~~&~~~~&~\\ 
~~...2&$^{~~}$&369&19.2&~176.32&~366.97&247&~42.8&1.32&~0.43&-105&~\\ 
16...1&$^{E2}$&~69&19.7&-250.18&-190.67&154&~32.0&0.17&~~~~~&~~~~&~\\ 
~~...2&$^{E2}$&116&19.7&-250.07&-190.95&265&~54.3&0.21&~2.34&~-68&~\\ 
17...1&$^{~~}$&177&19.9&~172.04&~407.95&311&~25.4&1.28&~~~~~&~~~~&~\\ 
~~...2&$^{~~}$&309&20.1&~171.76&~407.83&364&~33.4&2.18&~0.82&-156&~\\ 
18...1&$^{E2}$&~69&19.9&-250.13&-190.70&187&~38.7&0.25&~~~~~&~~~~&~\\ 
~~...2&$^{E2}$&116&20.1&-250.07&-190.91&326&~66.6&0.60&~1.70&~-75&f\\ 
19...1&$^{~~}$&~~0&19.9&~179.71&~352.89&614&~44.8&0.85&~~~~~&~~~~&~\\ 
~~...2&$^{~~}$&~69&19.9&~179.93&~352.83&561&116.4&0.68&~1.18&~-16&~\\ 
~~...3&$^{~~}$&177&20.1&~180.20&~352.83&810&~66.3&0.79&~1.03&~~-7&g\\ 
~~...4&$^{~~}$&369&19.5&~180.99&~350.57&676&116.9&1.92&~2.62&~-61&~\\ 
20...1&$^{C~}$&~69&19.9&~181.82&~370.31&168&~34.8&0.65&~~~~~&~~~~&~\\ 
~~...2&$^{C~}$&309&19.9&~181.82&~370.27&293&~26.9&0.50&~0.07&~-83&~\\ 
~~...3&$^{C~}$&369&19.9&~181.93&~370.17&192&~33.3&1.33&~0.23&~-52&~\\ 
21...1&$^{E2}$&~69&20.1&-250.18&-190.70&159&~32.9&0.20&~~~~~&~~~~&~\\ 
~~...2&$^{E2}$&116&20.1&-250.07&-190.91&326&~66.6&0.60&~1.86&~-62&f\\ 
22...1&$^{~~}$&~~0&20.1&~179.77&~352.91&471&~34.3&1.14&~~~~~&~~~~&~\\ 
~~...2&$^{~~}$&~69&20.1&~179.98&~352.86&339&~70.2&0.86&~1.16&~-11&~\\ 
~~...3&$^{~~}$&177&20.1&~180.20&~352.83&810&~66.3&0.79&~0.92&~~-9&g\\ \hline
23...1&$^{~~}$&177&20.1&~223.10&-608.20&~86&~~7.0&0.10&~~~~~&~~~~&~\\ 
~~...2&$^{~~}$&369&20.1&~223.44&-609.30&~57&~~9.9&1.80&~2.21&~-73&~\\ 
24...1&$^{F1}$&116&20.1&~229.32&-685.79&~39&~~8.0&0.29&~~~~~&~~~~&~\\ 
~~...2&$^{F1}$&369&19.9&~228.88&-685.77&~46&~~7.9&1.42&~0.64&~177&~\\ 
25...1&$^{F2}$&116&20.1&~227.88&-654.77&~37&~~7.6&0.75&~~~~~&~~~~&~\\ 
~~...2&$^{F2}$&369&19.9&~228.60&-654.33&~49&~~8.5&1.06&~1.23&~~31&~\\ 
26...1&$^{F3}$&116&20.1&~220.82&-602.79&~40&~~8.1&0.76&~~~~~&~~~~&~\\ 
~~...2&$^{F3}$&369&19.9&~221.77&-601.88&~50&~~8.7&0.93&~1.90&~~44&~\\ \hline 
27...1&$^{G~}$&309&18.2&-967.49&-380.66&119&~10.9&0.47&~~~~~&~~~~&~\\ 
~~...2&$^{G~}$&369&18.2&-967.49&-380.82&~77&~13.3&0.30&~0.96&~-90&~\\ \hline
\enddata  
\tablecomments {{\footnotesize
Column 1: Detection of proper motion. Subscription indicates the affiliation to the maser cluster.
Column 2: Observing day,
Column 3: Velocity of the spot,
Column 4: $\alpha$ position of the spot.
Column 5: $\delta$ position of the spot. The coordinate system is the same as in Figure \ref{fig:pmarrow}.
Column 6: Intensity in Jansky.
Column 7: Signal to noise ratio of the spot.
Column 8: Size of the maser spot(HPBW). 
Column 9: Dimension of the proper motion.
Column 10: Position angle of the direction of the proper motion.
Column 11: Overlap selection of the maser spot selection.    }}
\label{tab:PMT2}
\end{deluxetable}
\end{landscape}

\begin{deluxetable}{lrrrrrrrrrrl}
\tablecolumns{12} \tablewidth{0pc}
\tablecaption{Relative Proper Motions of Independent Clusters.}
\tablehead {
\multicolumn{1}{c}{\footnotesize{Cluster}}&\multicolumn{1}{c}{\footnotesize{Velocity}}&\multicolumn{2}{c}{\footnotesize{Relative Position}}&\multicolumn{2}{c}{\footnotesize{Intensity}}&\multicolumn{3}{c}{\footnotesize{Proper Motion}}\\
 & \footnotesize{\vlsr} & \footnotesize{$\alpha$} & \footnotesize{$\delta$} &&&\scriptsize{$\mu_{\alpha}$} &
 \footnotesize{$\mu_{\delta}$} & \footnotesize{$\mu$} &  \\
 &\scriptsize{(\kms)}&\scriptsize{(mas)}&\scriptsize{(mas)}&\scriptsize{(Jy)}&
  \scriptsize{\it(S.N.R.)} &\scriptsize{(mas~yr$^{-1}$)}&\scriptsize{(mas~yr$^{-1}$)}&\scriptsize{(mas~yr$^{-1}$)}&
}
\startdata
1.~D1 &  8.7& -18.96&  57.09&722& 98  & 0.68&-0.71& 0.98&~\\
2.~D2 & 10.2& -19.32&  57.06&383& 55  & 0.90&-0.62& 1.15&~\\
3.~E1 & 18.3&-234.72&-183.04&273& 56  & 0.67&-1.25& 1.43&~\\
4.~E2 & 19.9&-250.12&-190.80&236& 49  & 0.74&-1.81& 1.97&~\\
5.~C & 19.9& 181.86& 370.25&218& 32  & 0.34&-0.36& 0.51&~\\
6.~F1 & 20.0& 229.10&-685.78& 43&  8  &-0.64& 0.03& 0.64&~\\
7.~F2 & 20.0& 228.24&-654.55& 43&  8  & 1.05& 0.64& 1.22&~\\
8.~F3 & 20.0& 221.30&-602.33& 45&  8  & 1.37& 1.31& 1.89&~\\
9.~G & 18.2&-967.49&-380.74& 98& 12  & 0.00&-0.96& 0.96&~\\ \hline
\enddata
\tablecomments { {\footnotesize
Column 1: Cluster Name,
Column 2: Velocity of the cluster,
Column 3: Mean $\alpha$ position of the cluster,
Column 4: Mean $\delta$ position of the cluster.
 The coordinate system is the same as in Figure \ref{fig:pm-igs}.
Column 7: Mean intensity in Jansky,
Column 8: Mean signal to noise ratio of the cluster,
Column 9: $\alpha$ component of the proper motion,
Column 10: $\delta$ component of the proper motion,
Column 11: Dimension of the proper motion,
}}
\label{tab:PMT}\end{deluxetable}

\begin{landscape}

\begin{table*}[htbp]
\caption{ \footnotesize 
Peculiar motions (deviation from a galactic circular rotation) of each maser cluster in S269. 
 $U$, $V$, and $W$ [\kms] are three components of peculiar motions. $U$ is toward the Galactic center,
 $V$ is toward Galactic rotation, and $W$ is toward the North Galactic Pole.
  Galactic radius and circular rotation of the LSR are assumed to be $R_0=8.0~{\rm kpc}$ and $\Theta_0=200~${\rm \kms}.   The flat rotation curve is assumed. 
  The solar motion relative to the LSR is and $(U_\odot, V_\odot, W_\odot)=(10.00, 5.25, 7.17)~$\kms \citep{DehnenBinney:98},}
\begin{tabular}{lrrrrrrrr}
    &&\multicolumn{2}{c}{relative to $S269_{H2007}$}&& \multicolumn{3}{c}{DB98}&  \\
\cline{3-4} \cline{6-8}
Cluster&\multicolumn{1}{c}{$v_{\rm LSR}$}&\multicolumn{1}{c}{$\mu_{\alpha}$}&\multicolumn{1}{c}{$\mu_{\delta}$}&&U~~&V~~&W~~&\\
        &\kms&mas~yr$^{-1}$&mas~y$r^{-1}$&&\kms&\kms&\kms&\\ \hline
$S269_{H2007}$&$19.60\pm0.25$&\multicolumn{1}{c}{\_}&\multicolumn{1}{c}{\_}&&$-0.1\pm~0.2$&$ -0.6\pm~0.2$&$~-4.5\pm~0.1$ \\ \hline
S269~C~&$19.9\pm0.0 $&$~0.34\pm0.47$&$-0.36\pm0.42$&&$~1.7\pm~0.8$&$-14.9\pm~5.6$&$~~9.4\pm10.3$ \\ 
S269~D1&$ 8.7\pm0.2 $&$~0.68\pm0.00$&$-0.71\pm0.03$&&$14.7\pm~0.2$&$-24.6\pm~0.2$&$~13.0\pm~0.1$ \\ 
S269~D2&$10.2\pm0.2 $&$~0.90\pm0.70$&$-0.62\pm0.18$&&$13.1\pm~1.2$&$-25.5\pm~8.3$&$~18.8\pm15.4$ \\ 
S269~E1&$18.3\pm0.1 $&$~0.67\pm0.31$&$-1.25\pm0.00$&&$~7.4\pm~0.5$&$-37.8\pm~3.7$&$~~6.0\pm~6.8$ \\ 
S269~E2&$19.9\pm0.2 $&$~0.74\pm0.25$&$-1.81\pm0.30$&&$~8.3\pm~0.5$&$-51.0\pm~3.0$&$~~0.8\pm~5.5$ \\ 
S269~F1&$20.0\pm0.2 $&$-0.64~~~~~~~~$&$~0.03~~~~~~~$&&$-1.4\pm~0.2$&$~~5.1\pm~0.2$&$~-7.5\pm~0.1$ \\ 
S269~F2&$20.0\pm0.2 $&$~1.15~~~~~~~~$&$~0.64~~~~~~~$&&$-1.4\pm~0.2$&$~-2.9\pm~0.2$&$~39.1\pm~0.1$ \\ 
S269~F3&$20.0\pm0.0 $&$~1.37~~~~~~~~$&$~1.31~~~~~~~$&&$-3.8~~~~~~~~$&$~~8.9~~~~~~~~$&$51.9~~~~~~~$ \\ 
S269~G~&$18.2\pm0.0 $&$~0.00~~~~~~~~$&$-0.96~~~~~~~$&&$~5.4~~~~~~~~$&$-23.5~~~~~~~~$&$-5.2~~~~~~~$ \\ \hline
average&             &              &              &&$~4.4\pm~6.1$&$~-16.7\pm18.4$&$~12.2\pm18.7$ \\ \hline   
\end{tabular}
\label{tbl:UVW2}
\end{table*}

\end{landscape}
\newpage

\appendix


\section{Closure Phase of the $\vlsr=19.5$~\kms channel}\label{sect:closure}
\begin{landscape}
\begin{figure*}[htbp]
\begin{center}
\includegraphics[scale=0.75]{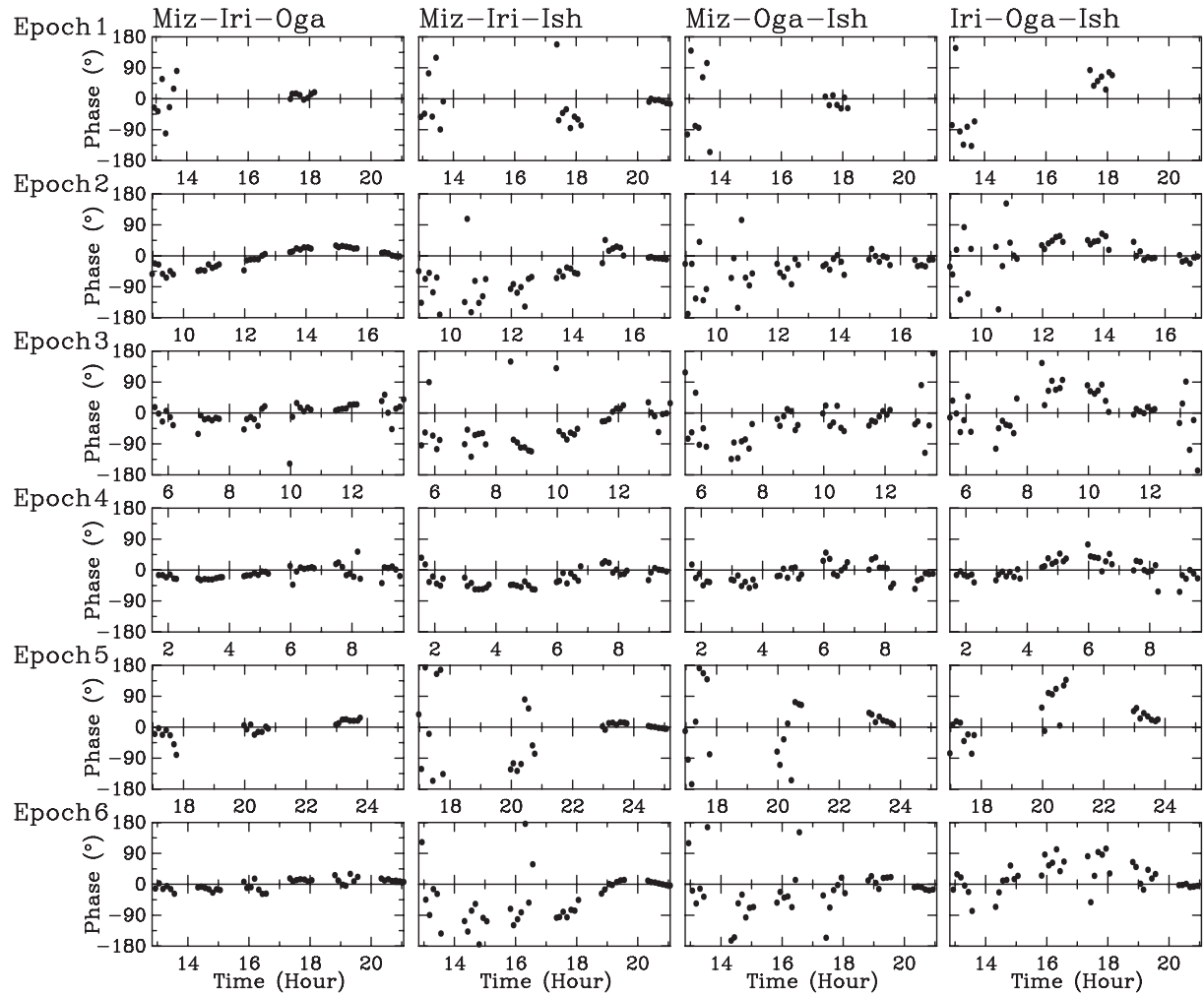}
\caption{Closure phases of the channel, $\vlsr=19.5$~\kms at the 6 observational epochs .
From left to right, those of the four triangles, Mizusawa-Iriki-Ogasawara, Mizusawa-Iriki-Ishigaki,
 Mizusawa-Ogasawara-Ishigaki, and Iriki-Ogasawara-Ishigaki are shown.
 Every closure phase is from 7.0 min integration.}
\label{fig:Closure}
\end{center}
\end{figure*}
\end{landscape}

To demonstrate the complex structure at the $\vlsr=19.5$~\kms channel
from visibility data, here we show the closure phases of the $\vlsr=19.5$~\kms channel, which
was used for the astrometry by H2007. We found non-zero phase values
in all the observational epochs (Figure~\ref{fig:Closure}).
 Non-zero closure phase indicates the existence of a complex structure.
 Closure phase $\Phi_{ABC}$ of a triangle composed of antennas A, B, and C is defined as follows.
\begin{eqnarray*}
 \Phi_{ABC}& \equiv  &\theta_{AB}^{obs} +\theta_{BC}^{obs}+\theta_{CA}^{obs}
\end{eqnarray*}
where, 
$\theta_{AB}^{obs}=\theta_{AB}+(\phi_{A}-\phi_{B})$, 
$\theta_{BC}^{obs}=\theta_{BC}+(\phi_{B}-\phi_{C})$ , and 
$\theta_{CA}^{obs}=\theta_{CA}+(\phi_{C}-\phi_{A})$.
$\theta_{XY}^{obs}$ is the observed fringe phase of the baseline between stations X and Y.
$\phi_{X}$ is antenna based phase error. $\theta_{XY}$ is the intrinsic phase due to the observed source structure.
If we substitute these in the equation,
\begin{eqnarray*}
 \Phi_{ABC} &=& \theta_{AB}^{obs} +\theta_{BC}^{obs}+\theta_{CA}^{obs} \\
            &=& \theta_{AB}+(\phi_{A}-\phi_{B}) \\
            &+& \theta_{BC}+(\phi_{B}-\phi_{C}) \\
            &+& \theta_{CA}+(\phi_{C}-\phi_{A}) \\
            &=& \theta_{AB}+\theta_{BC}+\theta_{CA} 
\end{eqnarray*}
 In the closure phase $\Phi_{ABC}$, the antenna based phase errors are canceled,
and the value of $\Phi_{ABC}$ is defined only by the phases due to the structure of the observed source.
(Noted that a baseline based error, if such exists, is not canceled.) 
The closure phase is an observable quantity, totally free from instrumental delay or phase error
 at respective antennas, and is defined only by the structure of the observed source.
The closure phase is very sensitive to asymmetry of the structure.
A source whose structure is point symmetric yields zero closure phases for any array geometry.
Conversely, the closure phase of an asymmetric source is non-zero \citep{Jennison:58, Pearson:84}.

 None of the structures of the \hho masers at the $\vlsr=19.5$~\kms
channel were any kind of point symmetric structure including a single point. 
These closure phases also show time variations between the observational epochs, 
reflecting the structure changes in the velocity channel.

\end{document}